\documentclass[conference]{IEEEtran}
\IEEEoverridecommandlockouts
\usepackage{cite}
\usepackage{amsmath,amssymb,amsfonts}
\usepackage{algorithmic}
\usepackage{graphicx}
\usepackage{float}
\usepackage{subfigure}
\usepackage{textcomp}
\usepackage{xcolor}
\def\BibTeX{{\rm B\kern-.05em{\sc i\kern-.025em b}\kern-.08em
    T\kern-.1667em\lower.7ex\hbox{E}\kern-.125emX}}
\begin{document}

\title{Propagation Path Loss Models in Forest Scenario at 605 MHz\\

\thanks{This work was supported in part by the National Key Research and Development Program of China under Grant No. 2020YFC1511801, in part by the Beijing Municipal Natural Science Foundation under Grant L192030, and in part by the National Natural Science Foundation of China (NSFC) under Grants No. 62171054, and No. 61871416.}
}


\author{
\IEEEauthorblockN{
Zhe Xiao\textsuperscript{1},  Shu Sun\textsuperscript{2}, Zhenyu Liu\textsuperscript{1}, Lianming Xu\textsuperscript{3},  Wei Huang\textsuperscript{4}, Li Wang\textsuperscript{1}, and Aiguo Fei\textsuperscript{1}}
\IEEEauthorblockA{
\textit{1.~School of Computer Science (National Pilot Software Engineering School), Beijing University of Posts and}\\
\textit{Telecommunications, Beijing, China}\\
\textit{2.~Department of Electronic Engineering, School of Electronic Information and Electrical Engineering, Shanghai }\\
\textit{Jiao Tong University, Shanghai, China} \\
\textit{3.~School of Electronic Engineering, Beijing University of Posts and Telecommunications, Beijing, China}\\
\textit{4.~Chengdu TD Tech Ltd, Chengdu, China}\\
Email:
\{xiaozhe, lzyu, xulianming, liwang, aiguofei\}@bupt.edu.cn;\\ shusun@sjtu.edu.cn;\\ wei.huang@td-tech.com}
}

\maketitle

\begin{abstract}

When signals propagate through forest areas, they will be affected by environmental factors such as vegetation. Different types of environments have different influences on signal attenuation. This paper analyzes the existing classical propagation path loss models and the model with excess loss caused by forest areas and then proposes a new short-range wireless channel propagation model, which can be applied to different types of forest environments. We conducted continuous-wave measurements at a center frequency of 605 MHz on predetermined routes in distinct types of forest areas and recorded the reference signal received power. Then, we use various path loss models to fit the measured data based on different vegetation types and distributions. Simulation results show that the proposed model has substantially smaller fitting errors with reasonable computational complexity, as compared with representative traditional counterparts.
\end{abstract}


\section{Introduction}
Different natural landforms such as valleys, forests, and other areas have a significant impact on the propagation of radio waves\cite{b1}. The forest area is covered by dense trees, which makes the multipath fading and non-line-of-sight channel very complex, and differs from the effects of urban buildings\cite{b7}. Some main factors affecting radio wave propagation, such as the distance between the transmitting and receiving antennas, the height of the antenna, and the type of ground objects, are reflected in the path loss formula as variable functions\cite{b8}. However, in different geographical environments, topographic relief, vegetation height and density, climate, and other factors have various degrees of influence on propagation\cite{b9}. Therefore, when these propagation models are applied in specific environments, the corresponding variable functions should be different, and it is necessary to identify a reasonable channel model. The accuracy of the channel model is very important for network deployment, because the inappropriate channel models will lead to significant reduction of network coverage\cite{b10}. 

Different typical propagation models have different characteristics, and they are applicable to different environment. Typical propagation models include Okumura Hata model\cite{b18}, cost-231 Hata model\cite{b19}, SPM model\cite{b20}, etc. These models are suitable for cities, suburbs and villages, but not for forest areas. In\cite{b16}, the authors proposed the Erceg model which is based on extensive experimental data collected by AT\&T Wireless Services across the United States in 95 existing macro cells at 1.9GHz. The terrains are classified in three categories. Category A is hilly terrain with moderate-to-heavy tree density, category B is hilly terrain with light tree density or flat terrain with moderate-to-heavy tree density and category C is mostly flat terrain with light tree density. Soon later, In\cite{b17}, Stanford University proposed Stanford University Interim (SUI) channel model is a set of 6 channel models representing three terrain types and a variety of Doppler spreads, delay spread and line-of-sight/non-line-of-site conditions that are typical of the continental US. The terrain type A, B, C are same as those defined earlier for Erceg model. However, these models were proposed earlier and are mainly applicable to the North American environment.

Other scholars have proposed channel models for specific forest environment, in addition to the  typical channel models. In \cite{b11}, the authors studied the oblique leaf path of roadside woodland, including three vegetation types, and obtained the attenuation loss results of oblique path at C-band yielding 0.9 dB overall improvement and up to 20 dB regional improvement in root mean square errors. In \cite{b12}, the authors investigated the propagation behavior of 28-GHz millimeter wave in coniferous forests and model its basic transmission loss, and proposed novel fully automated site-specific models. The root-mean-square deviations between model predictions and simulation results are 11.3 dB for an ITU woodland model and 6.8 dB for a site-specific model published in this paper. In \cite{b13}, the authors characterized the wireless channel for a short-range, temperate, medium density forest environment at the 5-GHz band. In \cite{b14}, the authors presented measurement results and propose empirical models of ultra-wideband signal propagation in forest environments after recording more than 22000 measurements at 165 sites in four different forest environments in Virginia and Maryland. However, these works only focus on the single scenario.

In this paper, we carry out propagation measurement campaigns in two different types of forest areas, and  record the measured values of signal propagation loss in those areas. Owing to the large attenuation of the signal in the forest area, the propagation effect of the lower frequency band is better, and in order to study the communication status of the emergency frequency band, so we adopted the 605 MHz frequency band for measurement. Then we use three classical large-scale path loss models and forest excess attenuation models to characterize the measured data, providing a comprehensive model comparison. Through the analysis of the results, we develop a new forest-specific path loss model, which has better performance than representative existing models.

\section{Description of measurements}


In order to study the impact of different forest areas on signal propagation, we selected two different types of areas which are Jiaozi snow mountain and Pudu-river dry-hot valley where channel propagation measurement campaigns were conducted in March 2022.

\begin{figure}[htbp]
\centerline{\includegraphics[width=8.5cm]{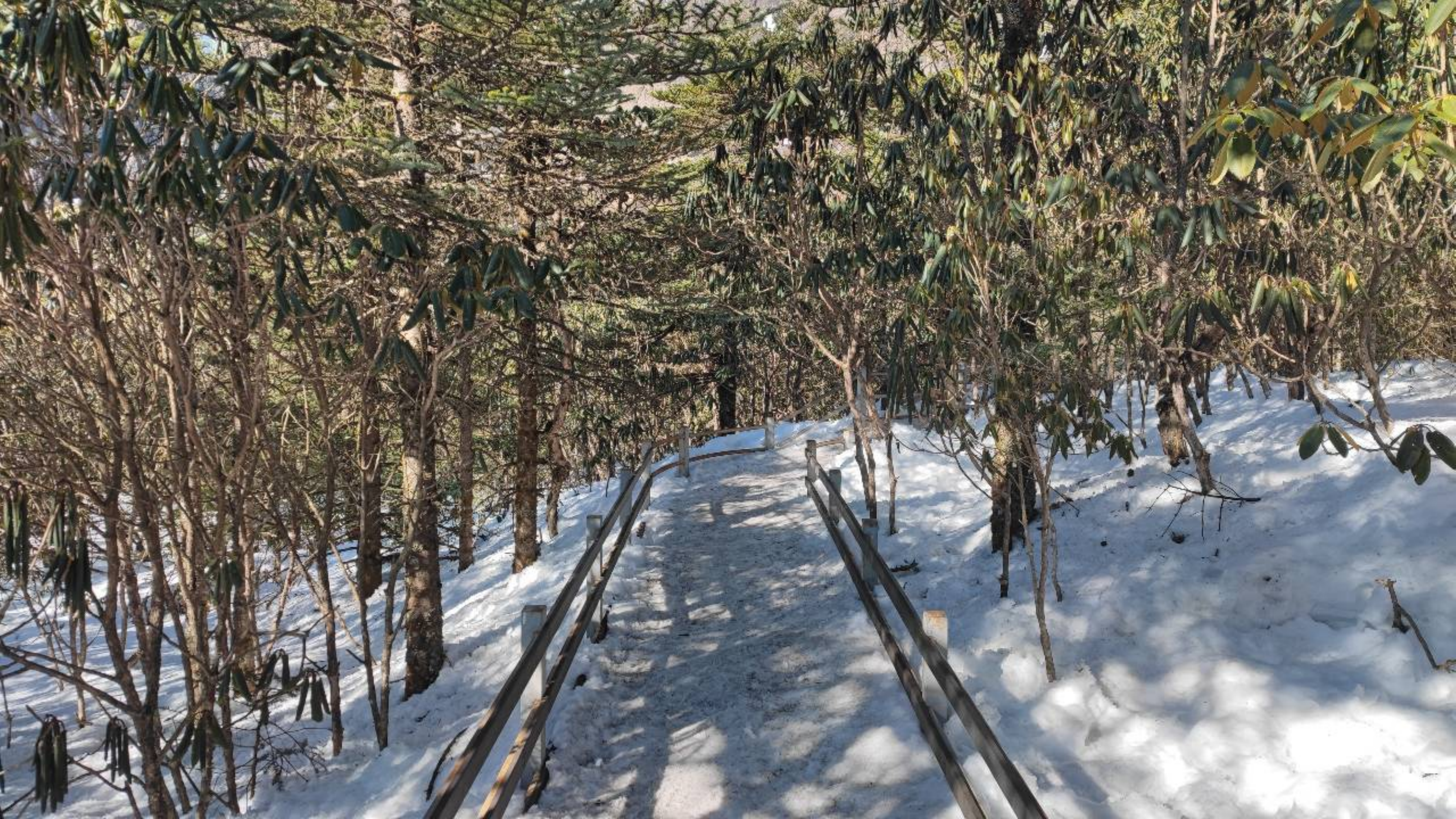}}
\caption{Geographical environment of Jiaozi snow mountain}
\label{fig}
\end{figure}

Jiaozi snow mountain is located at the junction of Luquan County and Dongchuan District in Yunnan Province, China, with a maximum altitude of 4344.1 meters (m), a minimum altitude of 2300 m, and a relative height difference of more than 2000 m. Jiaozi snow mountain belongs to seasonal snow mountain, which is the lowest snow mountain in the northern hemisphere. There are 15000 mu of Abies lanceolata primary secondary forest and Rhododendron forest, which are typical dense forest scenes. Fig. 1 shows the geographical scenario of Jiaozi snow mountain.

Pudu-river dry-hot valley is located in the Pudu river area in Yunnan Province as well. There are seven vegetation types, 11 vegetation subtypes, 17 formation groups, and 28 formations in the reserve, including dry-hot valley hard leaf evergreen oak forest, semi-humid evergreen broad-leaved forest, mountaintop bryophyte dwarf forest, cold temperate shrub, and cold temperate meadow. Fig. 2 shows the geographical scenario of Pudu-river dry-hot valley.


\begin{figure}[htbp]
\centerline{\includegraphics[width=8.5cm]{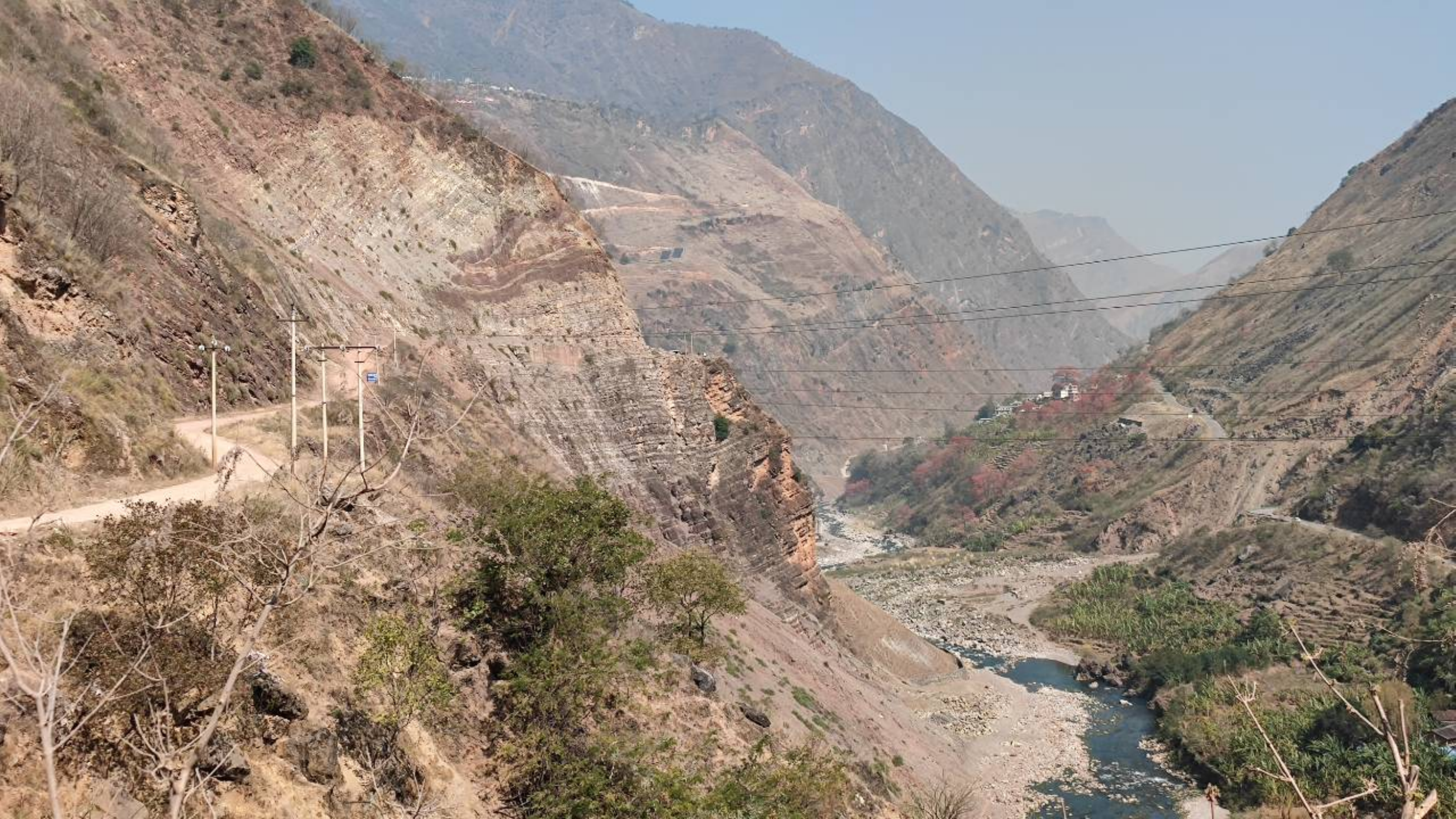}}
\caption{Geographical environment of Pudu-river dry-hot valley.}
\label{fig}
\end{figure}

\begin{figure}[htbp]
\centerline{\includegraphics[width=8.5cm]{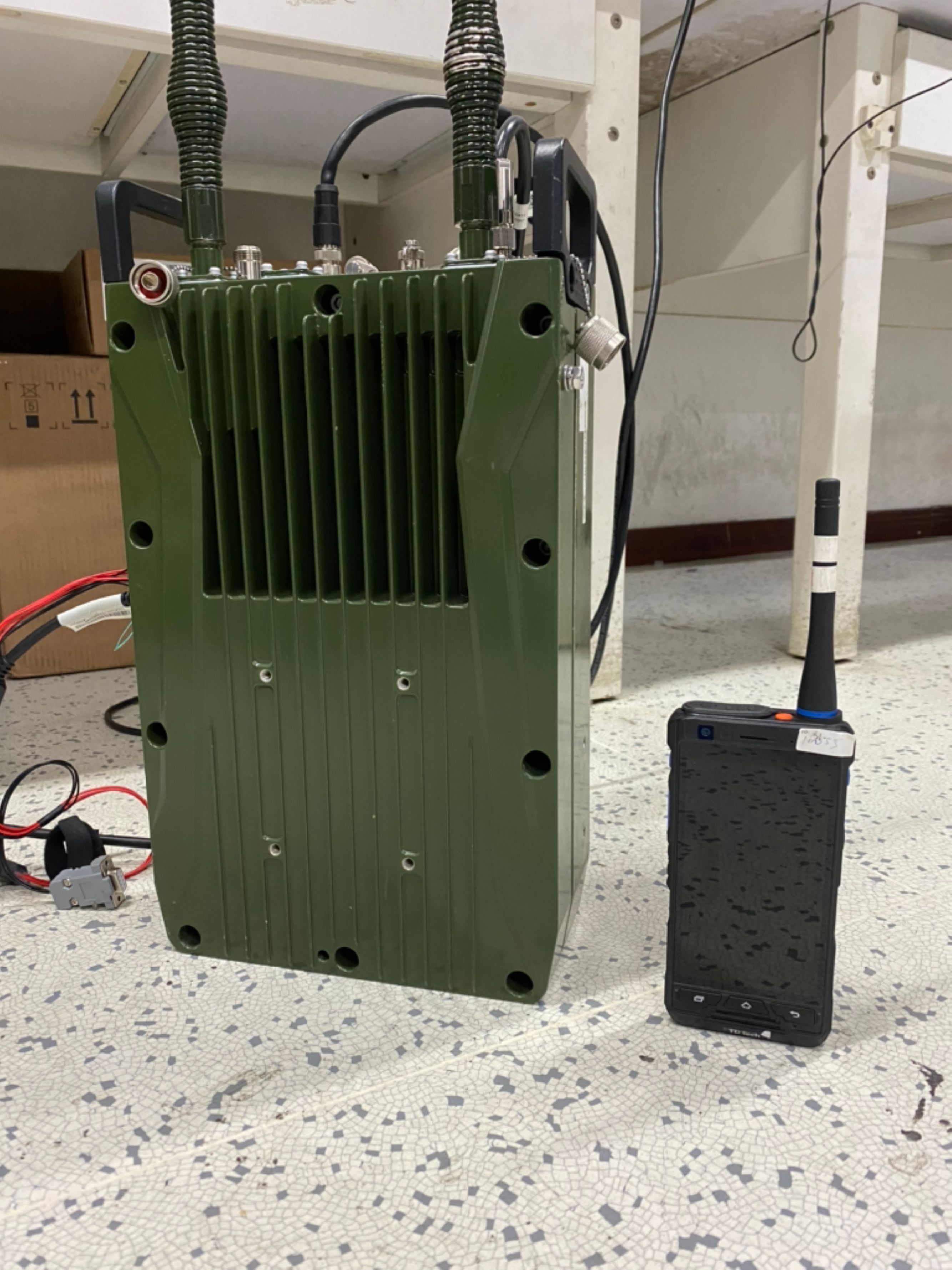}}
\caption{Transmitter and receiver equipment. The backpack base station  on the left is transmitter and the handheld device on the right is receiver.}
\label{fig}
\end{figure}

\begin{table}[htbp]
\caption{MEASUREMENT INFORMATION}
\begin{center}
\begin{tabular}{|l|l|}
\hline
Base station information                & Parameters or information  \\ \hline
Transmitting antenna height             & 1.5 m                       \\ \hline
Coordinate of Jiaozi snow mountain      & (102.848226, 26.0845327)   \\ \hline
Coordinate of Pudu-river dry-hot valley & (102.7342136, 26.02112129) \\ \hline
Base station transmission power         & 43 dBm                      \\ \hline
Base station antenna type               & Omnidirectional antennas   \\ \hline
Transmit antenna gain                   & 5 dBi                       \\ \hline
Receiving antenna gain                  & 0 dBi                       \\ \hline
Transmission band                       & 605 MHz                     \\ \hline
Cell reference signal power                               & 15.2 dBm                    \\ \hline                         
\end{tabular}
\end{center}
\end{table}

In the propagation measurement campaigns, we used a backpack base station as the signal transmitter (Tx) in a fixed position, equipped with an omni-directional antenna, while a handheld device was used as the receiver (Rx), with an omni-directional receiving antenna inside. Tx and Rx devices are shown in Fig. 3. The transmit power of the base station is 43 dBm, the transmit antenna gain is 2.5 dBi, the receive antenna gain is 0 dBi, and the carrier frequency is 605 MHz. The relevant information of the measurements is provided in Table \uppercase\expandafter{\romannumeral1}. We recorded the longitude and latitude of every transmission and reception position, adopted the continuous wave as the signal source to transmit the signal, and carried out the on-board test on the preset routes. 

We collected and recorded the pilot signal received power at the on-board test cell phone. Since the test was aimed to obtain the path loss data in the actual network, the test data truly reflects the propagation of broadband signals in the local wireless environment. As there is no need to set up a base station by itself, this test scheme is simple and convenient. Fig. 4 illustrates the measurement trajectory, and different colors of trajectory indicate different reference signal received power where green represents the minimum and red represents the maximum.  The measurement data in Jiaozi snow mountain and Pudu-river dry-hot valley will be shown in Fig. 5.


\begin{figure}[t]
\centering  
\subfigure[]{
\label{Fig.sub.1}
\includegraphics[width=8cm,height = 5cm]{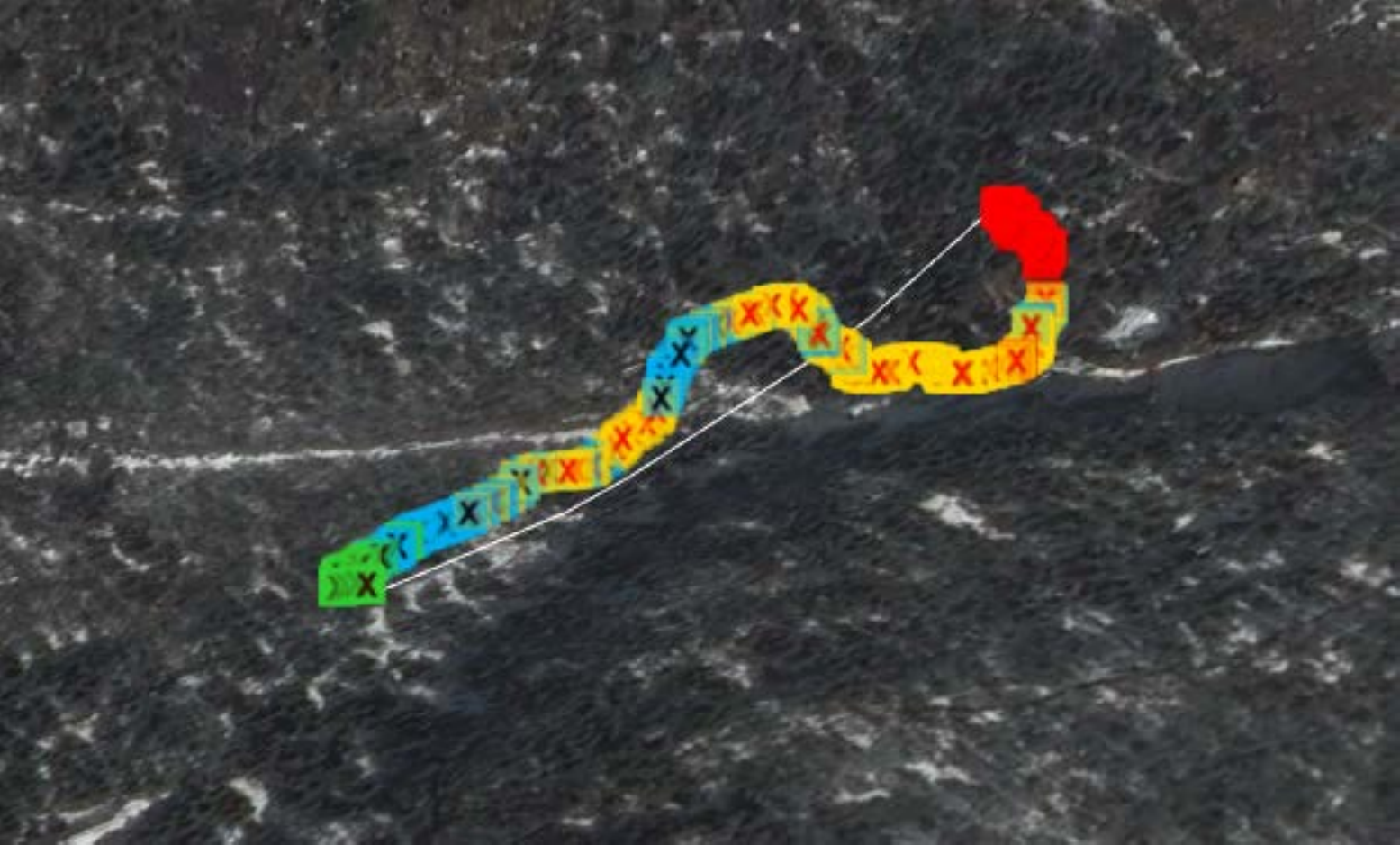}}\\
\subfigure[]{
\label{Fig.sub.2}
\includegraphics[width=8cm,height = 5cm]{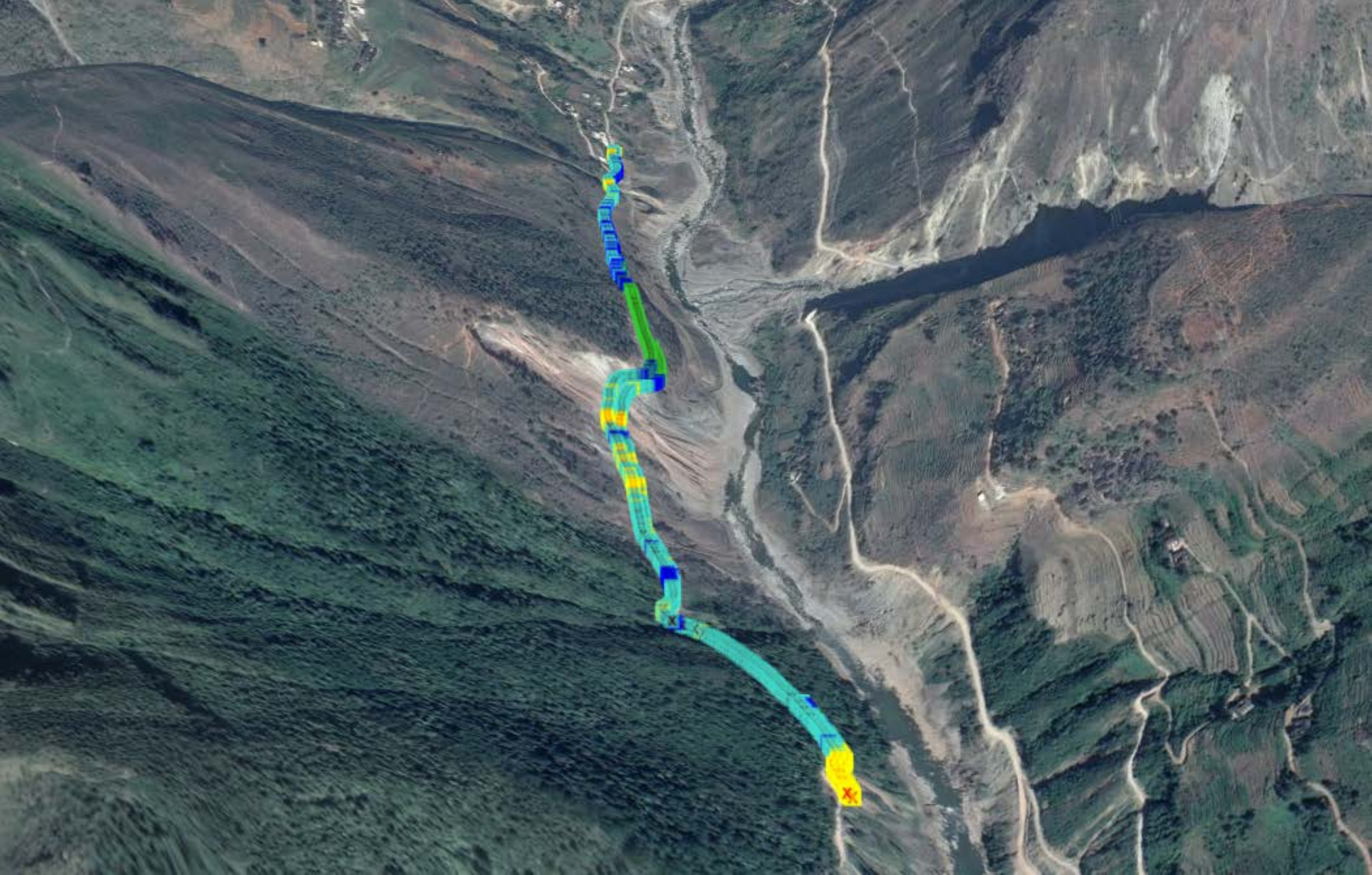}}
\caption{Measurement trajectory in Jiaozi snow mountain and Pudu-river Dry-hot valley. The colors in the graphs indicate reference signal received power value, where green and red represent the lowest and highest power, respectively.}
\label{1}
\end{figure}

\begin{figure}[t]
\centering  
\subfigure[]{
\label{Fig.sub.1}
\includegraphics[width=9cm]{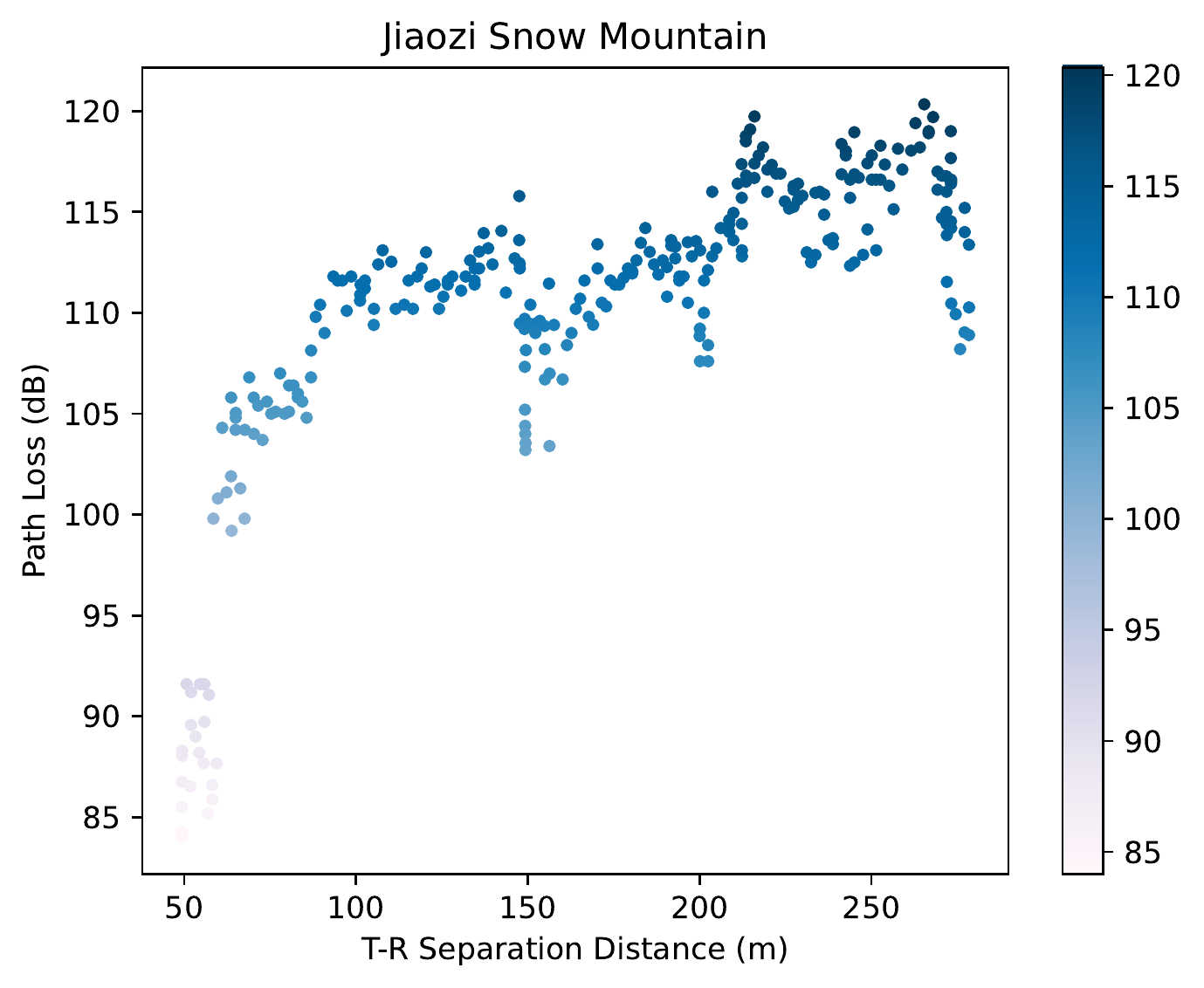}}\\
\subfigure[]{
\label{Fig.sub.2}
\includegraphics[width=9cm]{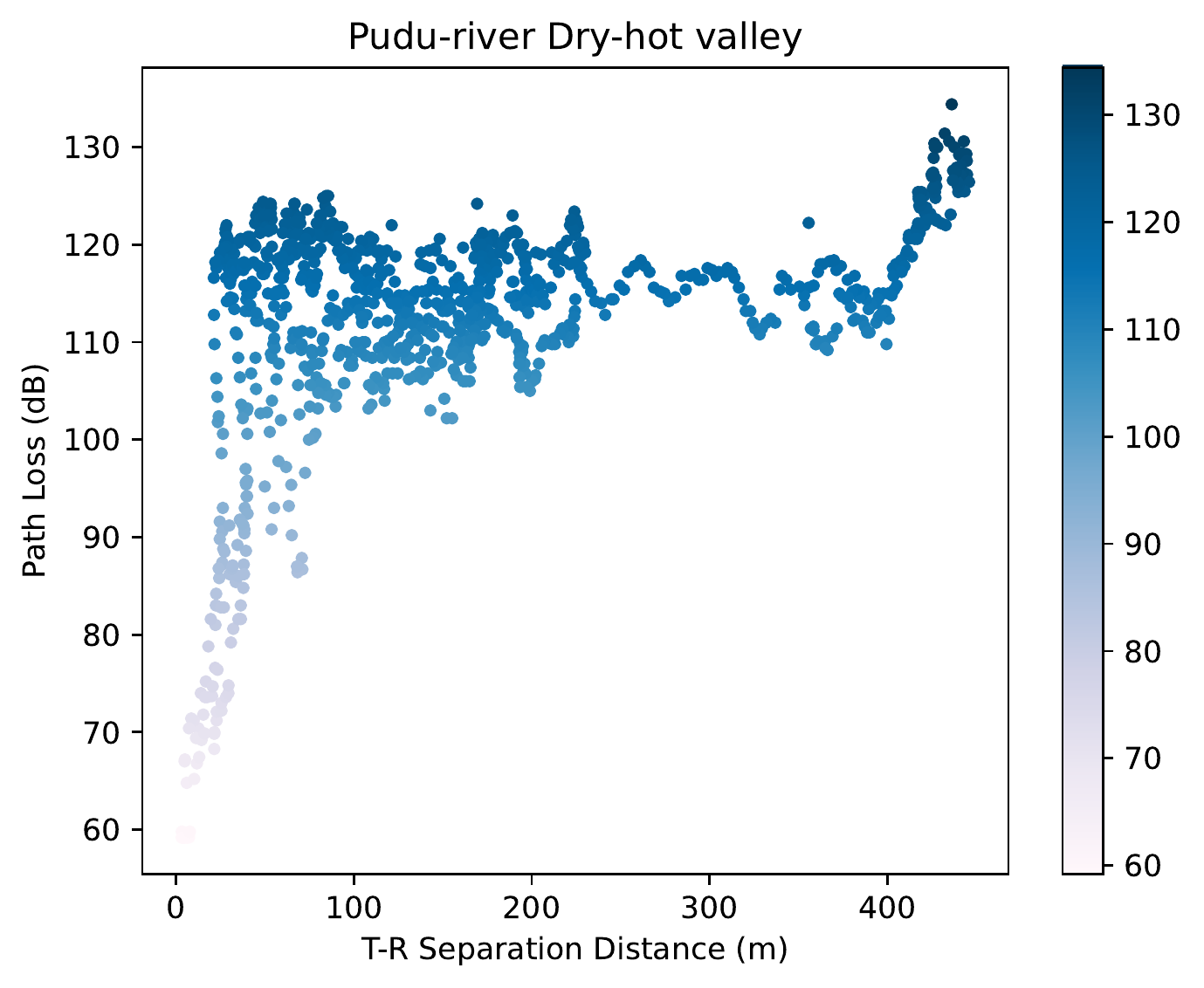}}
\caption{Measurement data set in Jiaozi snow mountain and Pudu-river dry-hot valley.}
\label{1}
\end{figure}

\section{Experimental results}

After obtaining the data set, we consider three path loss models as the baseline generic model which are the alpha-beta–gamma (ABG) model\cite{b2}\cite{b6}, the close-in free-space reference distance (CI) model\cite{b2}\cite{b6}\cite{b3}, and the free-space path loss (FSPL) model. The expressions of CI, ABG, and FSPL are as follows:

\begin{small}
\begin{equation}
\mathrm{P L}_{\mathrm{CI} }=10 n \log_{10}{\frac{d}{d_{0} } }+20\log_{10}{\left ( \frac{4\pi\times 10^{9}}{c}\right )}+20 \log_{10}{f} 
\end{equation}
\end{small}
\begin{small}
\begin{equation}
\mathrm{P L}_{\mathrm{ABG} }=10 \alpha\log_{10}{d} +\beta +10 \gamma\log_{10}{f}  
\end{equation}
\end{small}
\begin{small}
\begin{equation}
\mathrm{P L}_{\mathrm{FSPL} }=20\log_{10}{(\frac{4\pi fd\times 10^{9}}{c}  ) }
\end{equation}
\end{small}

\noindent where $n$ denotes the path loss exponent (PLE), $d_{0}$ is the close-in free-space reference distance and is set to 1 m\cite{b2}, $d$ is the 3-D T-R separation distance in meters, $\alpha$ and $\gamma$ are coefficients showing the dependence of path loss on distance and frequency, respectively, $\beta$ is an optimized offset value for path loss in decibels,  $f$ is the carrier frequency in GHz,  $c$ is the speed of light.

Note that the CI model has a very similar form compared with the ABG model, but has fewer model parameters and more solid physical basis\cite{b2}\cite{b6}. Since additional attenuation is caused by the occlusion of vegetation in the forest area, we use the ITU horizontal forest model\cite{b4} as the excess path loss model.The expressions of ITU horizontal forest model is as follows:

\begin{small}
\begin{equation}
\mathrm{P L}_{\mathrm{ITU-H} }=A_m\left[ 1-e^ {\left( -d\mu/A_m \right)} \right]
\end{equation}
\end{small}

\noindent where $\mu$ denotes the specific attenuation for very short vegetative paths (dB/m) and $A_m$ denotes the maximum attenuation for one terminal within a specific type and depth of vegetation (dB). Next, we combine FSPL model and ITU horizontal forest model to fit the measured data\cite{b5}. The expression of FSPL-H is given by:

\begin{small}
\begin{equation}
\mathrm{P L}_{\mathrm{FSPL-H} }=20\log_{10}{ (\frac{4\pi fd\times 10^{9}}{c}  )} +A_m\left[ 1-e^ {\left( -d\mu/A_m \right)} \right]
\end{equation}
\end{small}

\begin{figure}[t]
\centering  
\subfigure[]{
\label{Fig.sub.1}
\includegraphics[width=8.7cm]{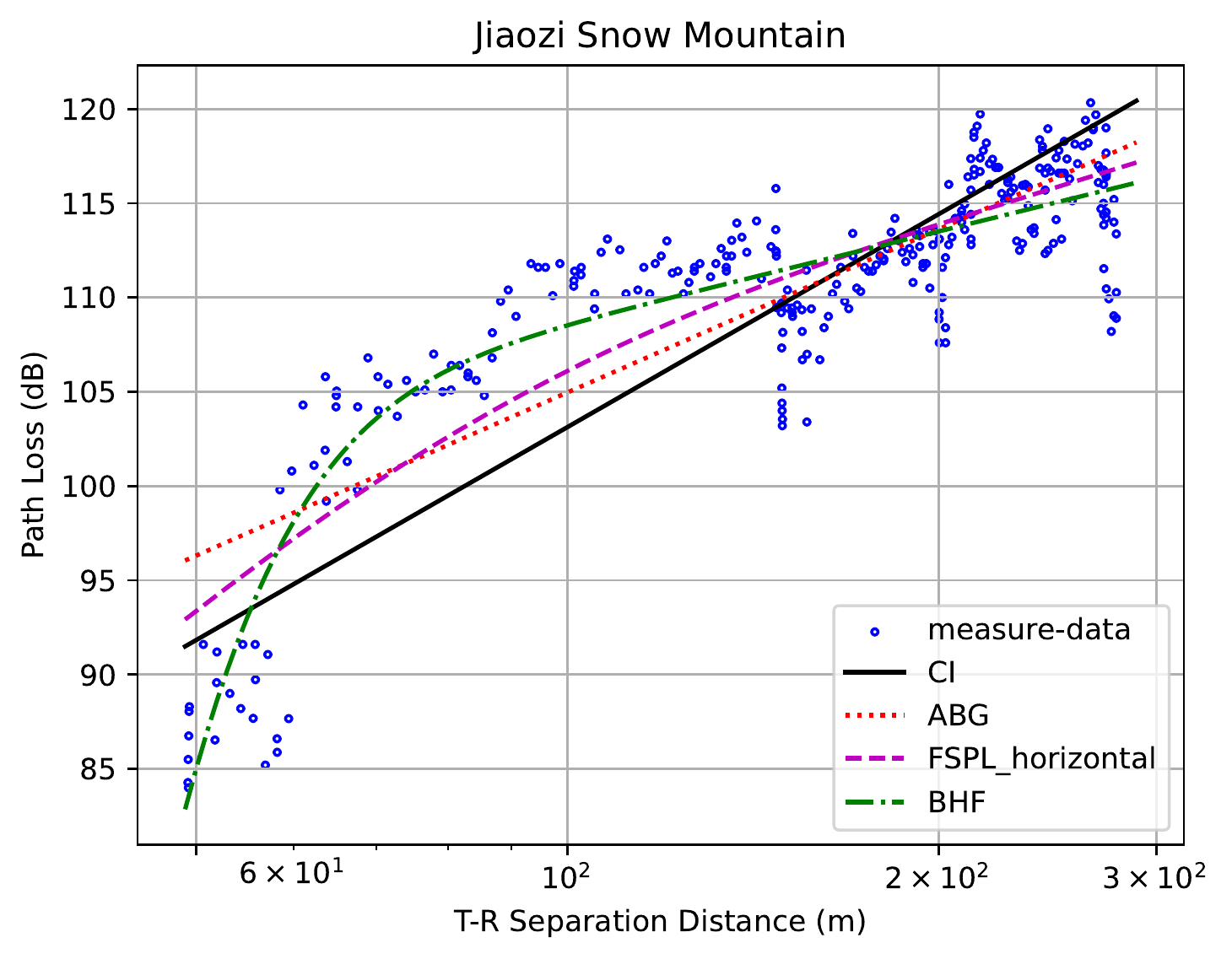}}\\
\subfigure[]{
\label{Fig.sub.2}
\includegraphics[width=8.5cm]{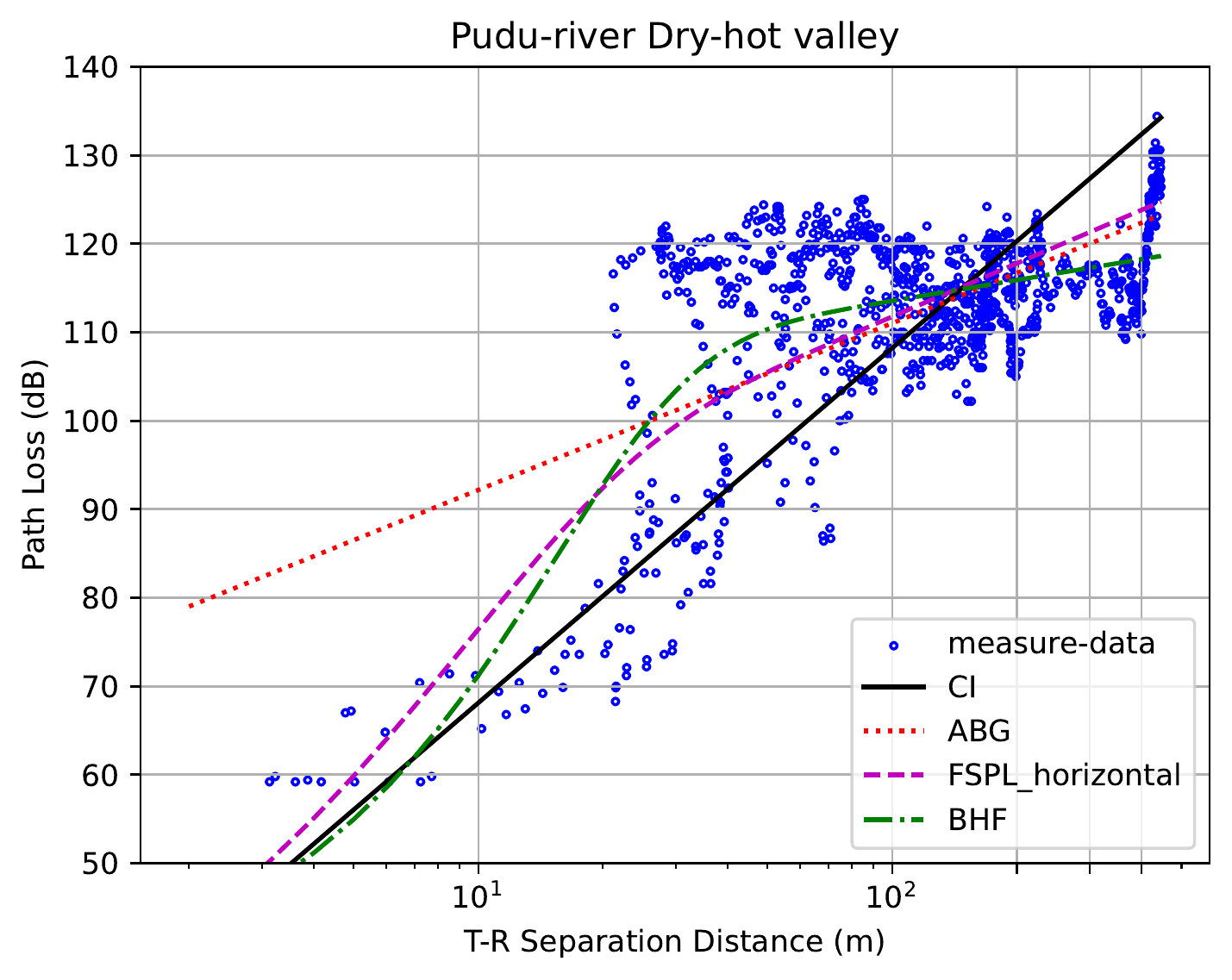}}
\caption{Measured path loss data and fitting results of the two classical models, the ITU forest excess attenuation model, and the proposed BHF model.}
\label{1}
\end{figure}

For short-distance specific forest scenes, we build a simple but powerful scene-specific model by more carefully characterizing forest-specific propagation loss, which can simplify the expression and parameters compared with directly combining two types of models presented above. We name the model Beijing University of Posts and Telecommunications horizontal forest model (BHF). The expression of BHF is as follows:

\begin{small}
\begin{equation}
\mathrm{P L}_{\mathrm{BHF}}=10\alpha  \log_{10}{d}+\beta+\zeta \tanh (d /20)+20 \log_{10}{f} 
\end{equation}
\end{small}

\noindent where $\alpha$ is a coefficient showing the dependence of path loss on the conventional log-scaled distance, $\beta$ is an optimized offset value for path loss in decibels, $\zeta$ is a coefficient characterizing the path loss caused by vegetation attenuation.

\begin{table}[htbp]
\caption{Optimized Model Parameters In The Baseline And Proposed Path Loss Models}
\begin{center}
\begin{tabular}{|l|l|l|l|l|}
\hline
Site                   & Model  & \begin{tabular}[c]{@{}l@{}}$n$(CI)\\$\alpha$(ABG)\\$Am$(FSPL-H) \\ $\alpha$(BHF)\end{tabular} & \begin{tabular}[c]{@{}l@{}}$\beta$(ABG)\\$\mu$(FSPL-H) \\$\beta$(BHF)\end{tabular} & \begin{tabular}[c]{@{}l@{}}$\gamma$(ABG)\\ $\zeta$(BHF)\end{tabular} \\ \hline
Jiaozi snow            & CI     & 3.8                                                   & -                                                  & -                                             \\ \cline{2-5} 
mountain               & ABG    & 2.9                                                   & 31.8                                              & 2.0                                           \\ \cline{2-5} 
\multicolumn{1}{|c|}{} & FSPL-H & 40.0                                                  & 1.2                                               & -                                              \\ \cline{2-5} 
                       & BHF    & 1.6                                                   & -1305.2                                           & 1407.0                                        \\ \hline
Pudu-river             & CI     & 4.0                                                   & -                                                  & -                                              \\ \cline{2-5} 
dry-hot valley         & ABG    & 1.9                                                   & 57.7                                              & 2.0                                           \\ \cline{2-5} 
                       & FSPL-H & 43.8                                                  & 4.6                                               & -                                              \\ \cline{2-5} 
                       & BHF    & 0.8                                                   & 48.3                                              & 64.2                                          \\ \hline
\end{tabular}
\end{center}
\end{table}

The BHF model contains the negative exponential characteristic attenuation caused by vegetation, where the changes are made on the basis of ITU horizontal model, and the additional attenuation is close to the hyperbolic tangent function. Compared with the ITU horizontal model in (4), the function about the additional attenuation of vegetation changes more gently in the BHF model.  We fit the BHF model and three baseline models by using the least-square method to find the optimal model parameters. The optimized model parameters are shown in Table  \uppercase\expandafter{\romannumeral2}. It can be seen from the Table \uppercase\expandafter{\romannumeral2} that the coefficients of the CI and FSPL-H models are relatively stable, with little change across environments, due to the free-space reference term acting as an anchor point. In contrast, the coefficients of the ABG and BHF models are strongly influenced by the environment.

Fig. 6 shows the fitting results of the CI, ABG, FSPL-H, and BHF models. It can be observed from Fig. 6 that the CI model and ABG model are two straight lines. Because the relationship between the parameter term and the 3D T-R separation distance in the expression is multiplication, when other coefficients are given, the function between path loss and distance in the logarithmic scale is a linear function. Since the adaptation ability of straight lines is relatively limited, the fitting errors of CI and ABG models are large. The CI model has fewer parameter variables than the ABG model, rendering larger errors. The relationship between $d$ and parameters of the other two models are more complex, so the fitting effect is better. It can be seen that the BHF model has stronger capability of adapting with the trend of measured data in Fig. 6(b).

\begin{table}[htbp]
\caption{Overall Models Performance}
\begin{center}
\begin{tabular}{|l|lll|l|}
\hline
-                           & \multicolumn{3}{l|}{Traditional}                                & Proposed \\ \hline
Model                      & \multicolumn{1}{l|}{CI}    & \multicolumn{1}{l|}{ABG}  & FSPL-H & BHF          \\ \hline
Jiaozi snow mountain       & \multicolumn{1}{l|}{4.6} & \multicolumn{1}{l|}{4.1} & 3.6   & 3.0         \\ \hline
Pudu-river dry-hot valley  & \multicolumn{1}{l|}{13.1}  & \multicolumn{1}{l|}{9.7} & 9.3   & 8.3         \\ \hline
Number of model parameters & \multicolumn{1}{l|}{1}     & \multicolumn{1}{l|}{2}    & 2      & 3            \\ \hline
\end{tabular}
\end{center}
\end{table}

We use root-mean-square error (RMSE) and the number of parameters to quantify the fitting effect of the models, which are given in Table \uppercase\expandafter{\romannumeral3}. Note that for single frequencies, $\gamma$ in the ABG model is set to 2, thus there are actually two model parameters in the ABG model.



It can be seen from Table \uppercase\expandafter{\romannumeral3} that the fitting error for Jiaozi snow mountain is smaller than that for Pudu-river dry-hot valley in general,  because the different types and densities of vegetation and the geographical environment have an impact on the transmission of signals. The terrain of Pudu-river dry-hot valley is steeper than that of  Jiaozi snow mountain. Accoding to Table \uppercase\expandafter{\romannumeral3},  the RMSEs of the BHF model are 3.0 dB and 8.3 dB for Jiaozi snow mountain and Pudu-river dry-hot valley, respectively. This shows that the BHF model proposed in this paper has the best fitting effect overall and is more suitable for the forest environment.

\section{Conclusion}

 In this paper, we have provided results from real-world measurement campaigns to assess channel characteristics for the forest environment. The signal measurement data near Jiaozi snow mountain and Pudu-river dry-hot valley are used to compare the attenuation of vegetation with comprehensive large-scale path loss models. Inspired by these results, we have developed a new site-specific path loss model. Compared with typical traditional models, the model proposed in this paper yields significantly smaller fitting errors with acceptable computational complexity, and is thus more suitable for the forest environment.


\vspace{12pt}

\end{document}